\documentclass[aps,prl,twocolumn,showpacs,a4paper]{revtex4-1}
\pdfoutput=1

\usepackage{geometry}
\usepackage{color}
\usepackage[latin1]{inputenc}
\usepackage{graphicx}
\usepackage{amsmath,amssymb}
\usepackage{afterpage}

\usepackage{textcomp}
\usepackage{amsfonts}
\usepackage{graphicx}
\usepackage{color}
\usepackage[latin1]{inputenc}
\usepackage{braket}
\usepackage{units}
\usepackage[svgnames]{xcolor}
\usepackage[
pdftitle={Site-resolved imaging of a fermionic Mott insulator},    
colorlinks=True,linkcolor=DarkSlateBlue,citecolor=DarkBlue,urlcolor=DarkBlue,
	pdfstartview=FitH,bookmarks=False,pdfpagemode=UseNone
]{hyperref}

\begin{document}
\title{Site-resolved imaging of a fermionic Mott insulator}

\date{\today}

\author{Daniel Greif}
\author{Maxwell F. Parsons}
\author{Anton Mazurenko}
\author{Christie S. Chiu}
\author{Sebastian Blatt}
\altaffiliation{Current address: Max-Planck-Institut f\"ur Quantenoptik, 85748 Garching, Germany}
\author{Florian Huber}
\author{Geoffrey Ji}
\author{Markus Greiner}
\email{greiner@physics.harvard.edu}
\affiliation{Department of Physics, Harvard University, Cambridge, Massachusetts 02138, USA}

\begin{abstract}
The complexity of quantum many-body systems originates from the interplay of strong interactions, quantum statistics, and the large number of quantum-mechanical degrees of freedom. Probing these systems on a microscopic level with single-site resolution offers important insights. Here we report site-resolved imaging of two-component fermionic Mott insulators, metals, and band insulators using ultracold atoms in a square lattice. For strong repulsive interactions we observe two-dimensional Mott insulators containing over 400 atoms. For intermediate interactions, we observe a coexistence of phases. From comparison to theory we find trap-averaged entropies per particle of $1.0\,k_{\mathrm{B}}$. In the band-insulator we find local entropies as low as $0.5\,k_{\mathrm{B}}$. Access to local observables will aid the understanding of fermionic many-body systems in regimes inaccessible by modern theoretical methods. 
\end{abstract}


\maketitle 

Detection and control of quantum many-body systems at the level of single lattice sites and single particles gives access to correlation functions and order parameters on a microscopic level. These capabilities promise insights into a number of complex and poorly understood quantum phases, such as spin-liquids, valence-bond solids, and $d$-wave superconductors \cite{Balents2010a, Anderson1987}. Site-resolved detection and control of the wave function remains elusive in conventional solid-state systems, but is possible in synthetic matter with ultracold atoms in optical lattices, because the associated length scales are accessible with high-resolution optical microscopy \cite{Nelson2007}. Site-resolved imaging of bosonic quantum gases has enabled studies of the superfluid-to-Mott insulator (MI) transition on a single-site level \cite{Bakr2010, Sherson2010a}, an experimental realization of Ising and Heisenberg spin chains \cite{Simon2011, Fukuhara2013a}, a direct measurement of entanglement entropy \cite{Islam2015}, and studies of dynamics of charge and spin degrees of freedom of quantum systems \cite{Fukuhara2013, Preiss2015}. The extension of site-resolved imaging to fermionic atoms may enable exploration of the rich phase diagram of the Hubbard model, thought to describe a variety of strongly-correlated quantum-mechanical phenomena, including high-temperature superconductivity \cite{Lee2006b}. Recently, site-resolved imaging of fermionic atoms has been demonstrated with different atomic species \cite{Haller2015, Cheuk2015, Parsons2015, Edge2015}, and a single-spin band insulator has been observed \cite{Omran2015}. Although fermionic Mott insulators and short-range anti-ferromagnetic spin correlations have been observed using conventional imaging techniques \cite{Jordens2008, Schneider2008a, Jordens2010, Taie2012b, Uehlinger2013a, Duarte2015, Messer2015, Greif2013, Hart2015a, Greif2015}, site-resolved imaging of interacting many-body states of fermions is an outstanding challenge.

\begin{figure*}[]
		\centering
    \includegraphics[width=\textwidth]{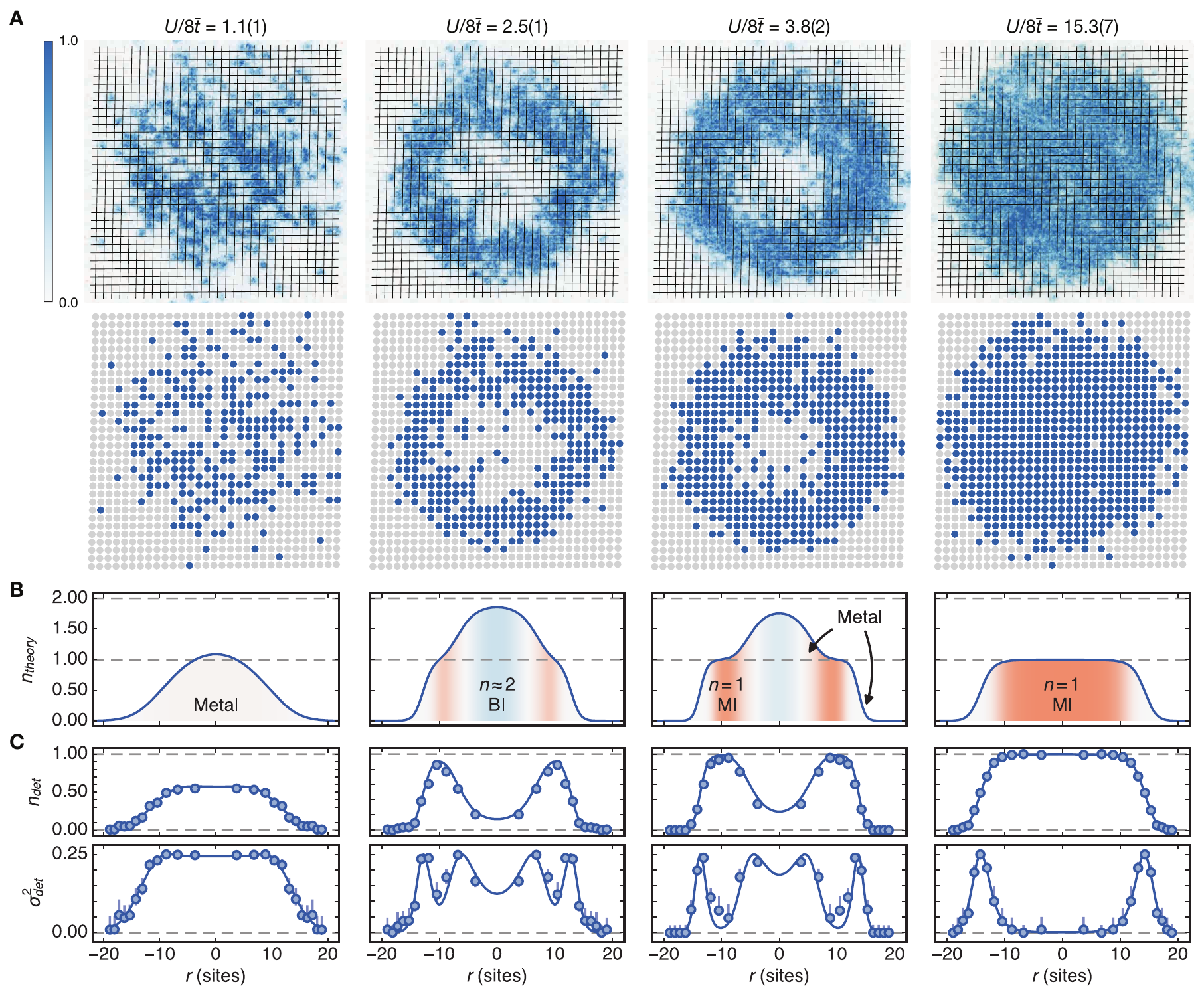}
\caption{{\bf Site-resolved images of the fermionic metal-to-Mott insulator transition.} (A) Single experimental images of the atoms in the square lattice are shown for varying interactions $U/8\overline{t}$, along with the extracted site occupation. Doublons are detected as empty sites because of light-assisted collisions during imaging. The color bar shows the normalized number of detected photons. 
    (B) Calculated full density profiles for the experimental parameters in (A) obtained by fits to the experimental data. Depending on the interaction we observe Mott-insulating (MI), band-insulating (BI), and metallic states. The MI appears for strong interactions as an extended spatial region with filling $n_{\mathrm{theory}}=1$. In the BI the filling approaches two. Metallic regions connect the different insulating states, where the filling changes by an integer. 
    (C) By applying azimuthal averages to the corresponding single images, we obtain radial profiles (mirrored at $r=0$) for the detected occupation $\overline{n_{\mathrm{det}}}$ and variance $\sigma^2_{\mathrm{det}}$. The variance is strongly reduced in the insulating phases. The radial profiles $\overline{n_{\mathrm{det}}}(r)$ were fit using a high-temperature series expansion of the single-band Hubbard model, with temperature and chemical potential as free parameters (solid lines). This gives temperatures $k_{\mathrm{B}}T/U = 1.6(3),\, 0.22(3),\, 0.12(2),$ and $0.050(7)$ from left to right, which are used for calculating the full density profiles in (B) and correspond to average entropies per particle of $S/N =2.2(3),\,1.00(8),\, 0.99(7),$ and $1.15(7)\,k_{\mathrm{B}}$. The large values of temperature and entropy at $U/8\overline{t}=1.1$ may be caused by non-adiabatic loading of the lattice at small interactions. Error bars are computed from a sampled Bernoulli distribution \cite{supplementary}.
}
	\label{fig1}
\end{figure*}

Here we demonstrate site-resolved observations of fermionic metals, MIs, and band insulators. A fermionic MI is a prime example of a strongly-interacting many-body state, where repulsive interactions between fermionic particles in two different spin states give rise to insulating behavior in a half-filled energy band. This behavior is well described by the Hubbard model. For low temperatures, a theoretical analysis of the phase diagram becomes difficult owing to the fermion sign problem \cite{Troyer2005}. At temperatures above the magnetic exchange energy, spin-order is absent, and a density-order crossover from a metallic state to a MI occurs when the ratio of interaction to kinetic energy is increased. Whereas the metallic state has a gapless excitation spectrum, is compressible, and shows a large variance in the site-resolved lattice occupation, the MI has a finite energy gap, is incompressible, and shows a vanishing variance in the occupation. Observing these characteristic properties of a MI requires temperatures well below the energy gap. For lower interactions and large filling an incompressible band insulator of doublons (two particles on a site) appears because of the Pauli exclusion principle. This behavior is in contrast to the bosonic case, where the absence of the Pauli exclusion principle allows higher fillings, and a ring structure of MI states with different integer fillings appears \cite{Bakr2010, Sherson2010a}. 

\begin{figure*}[tb]
\centering
    \includegraphics{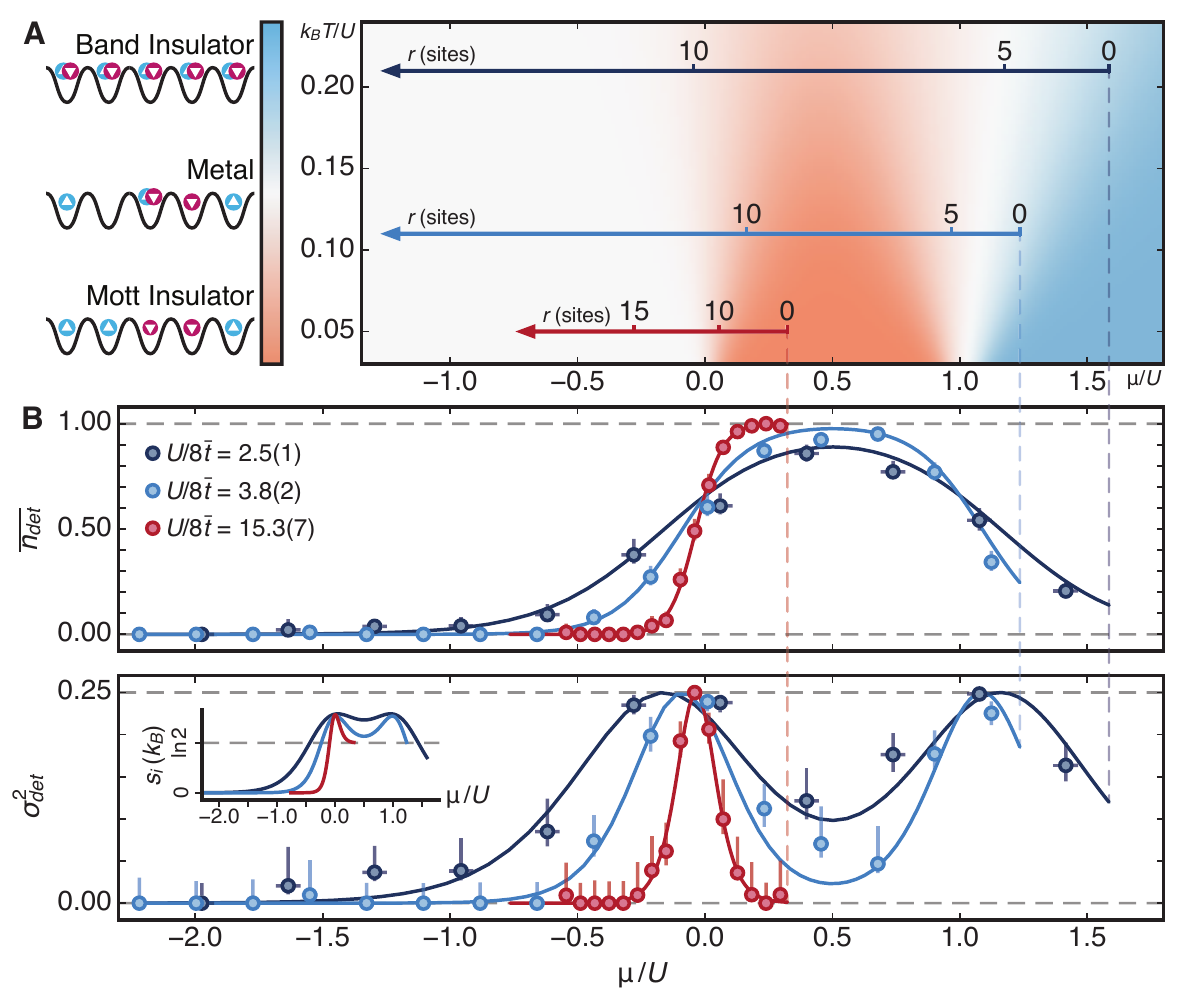}
    \caption{{\bf Cuts through the 2D Hubbard phase diagram.} (A) Schematic of the ($\mu / U$, $k_{\mathrm{B}}T/U$) phase diagram of the Hubbard model at $U \gg \overline{t}$, which is the relevant regime for all three data sets shown. The intensity of the shading reflects the normalized variance of the site occupation and the colors distinguish between MI and BI regimes. The arrows denote radial cuts through the trap for different interactions corresponding to the profiles shown in (B). (B) Detected site occupation and variance versus chemical potential, obtained from the radial profiles in Fig.\;1, with fits as solid lines \cite{supplementary} and error bars as in Fig. 1. The distributions are symmetric around $\mu=U/2$ (half-filling) from particle-hole symmetry.  (Inset, bottom left) Calculated local entropy per site $s_i$. }
	\label{fig2}
\end{figure*}

The starting point of the experiment is a low-temperature, two-dimensional gas of fermionic $^6\mathrm{Li}$ atoms with repulsive interactions in an equal mixture of the two lowest hyperfine ground states. After the preparation and cooling of the cloud, we set the $s$-wave scattering length to values $a=37\,a_0-515\,a_0$ by adjusting a magnetic bias field in the vicinity of the Feshbach resonance located at $832\,$G, where $a_0$ denotes the Bohr radius \cite{Zurn2013}. We then load the atoms into a square optical lattice using a $30$\,ms long linear ramp of the laser beam powers. The system is well described by a single-band two-dimensional Hubbard model on a square lattice with nearest-neighbor tunneling $t_x/h=0.279(10)$\,kHz and $t_y/h=0.133(4)$\,kHz along the two lattice directions with a corresponding bandwidth $8\overline{t}=1.65(7)$\,kHz and $U/h = 1.81(3)\, \mathrm{kHz}- 25.2(5)\,\mathrm{kHz}$. The explored ratios of $U/8\overline{t}$ thus range from the metallic ($U \sim 8\overline{t}$) to the MI ($U\gg 8\overline{t}$) regime. In the experiment an overall harmonic confinement is present with frequencies $\omega_x/2\pi=0.691(9)$\,kHz and $\omega_y/2\pi=0.604(3)$\,kHz. For more details see \cite{supplementary}. 

We detect the many-body state of the system by measuring the occupation of each lattice site with single-site resolution. For the detection we first rapidly increase all lattice depths to pin the atomic distribution and then image the fluorescence of the atoms with a high-resolution microscope onto an intensified CCD camera. Atoms on doubly-occupied sites are removed during the imaging as a consequence of light-assisted collisions \cite{DePue1999a}. We apply a deconvolution algorithm to the images to determine the occupation of every individual lattice site $n_{\mathrm{det}}$, which is unity for a single particle of either spin and zero for the case of an empty or doubly-occupied site. Our imaging technique allows a reliable determination of the site-resolved occupation, with an estimated imaging fidelity of $97.5(3)\%$ \cite{Parsons2015, supplementary}.

We directly observe the metal-to-MI transition on a site-resolved level. Single images show a drastic change in the occupation distribution when increasing the interaction, $U/8\overline{t}$, at constant atom number (Fig.\;1). For the weakest interactions ($U/8\overline{t} = 1.1(1)$) we observe a purely metallic state with a large occupation variance over the entire cloud. The maximum detected value for the variance is $0.25$, which is consistent with equal occupation probabilities of all four possible states per lattice site. The occupation decreases gradually for larger distances from the center because of the underlying harmonic confinement. In contrast, for the strongest interactions ($U/8\overline{t} = 15.3(7)$) we observe a large half-filled MI region containing about $400$ atoms \cite{supplementary}. The energy gap, $U$, suppresses the variance in occupation in the center of the cloud to values below $0.02$, with thermal excitations appearing as an increased variance in the occupation at the edge of the cloud. 

\begin{figure*}[tb]
\centering
	\includegraphics{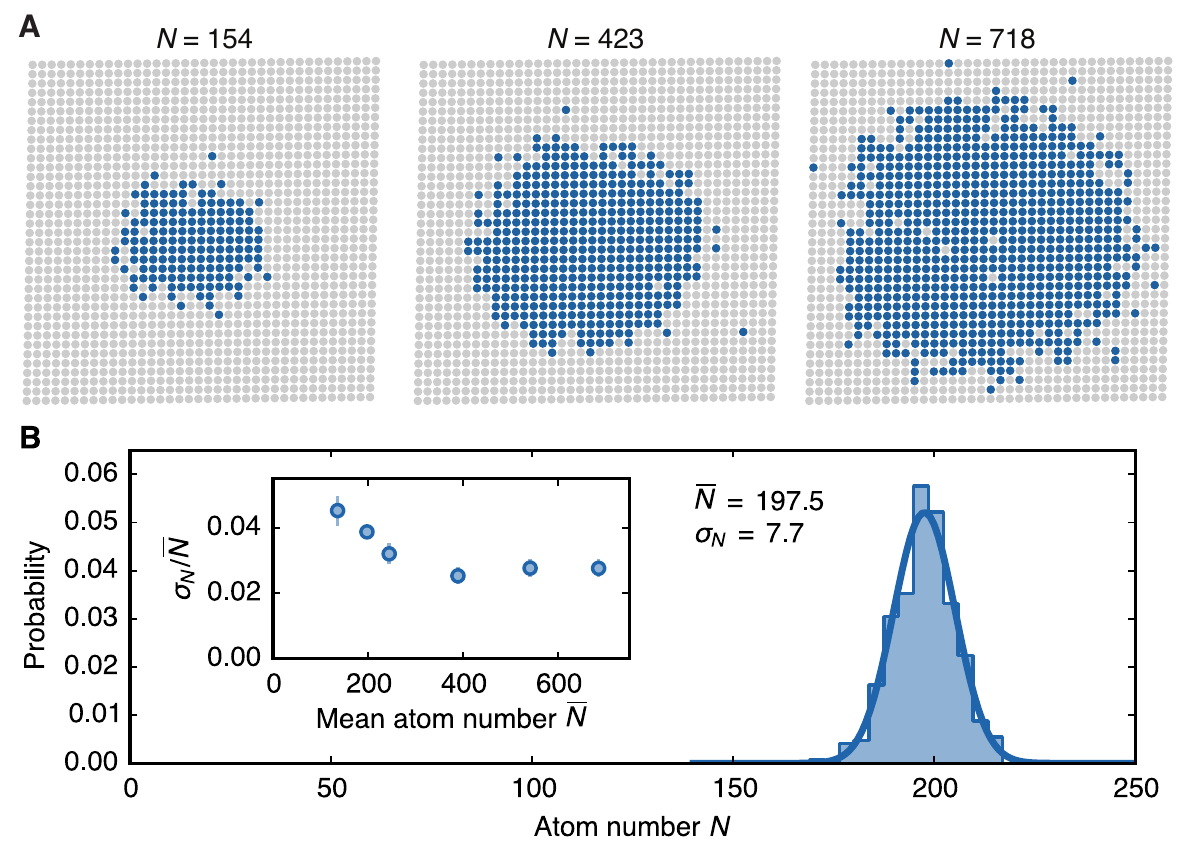}
    \caption{{\bf Controlling the size of a Mott insulator} (A) Single images of MI with varying detected atom numbers $N$ (without doublons) for $U/8\overline{t} = 15.3(7)$. (B) Histogram of detected atom numbers after $400$ experiment repetitions with the average atom number $\overline{N}$ and standard deviation $\sigma_N$ ($68\%$ confidence). The solid line shows the Gaussian distribution for the obtained $\overline{N}$ and $\sigma_N$. The inset shows the experimental stability of preparing samples with fixed atom number, characterized by the relative standard deviation for different $\overline{N}$ obtained from $\geq 50$ images. Error bars denote the respective standard errors. 
    }
	\label{fig3}
\end{figure*}

The underlying harmonic trap causes a spatially varying chemical potential, which can lead to the appearance of different phases within the same atomic cloud. For intermediate interactions ($U/8\overline{t}=2.5(1)$ and $U/8\overline{t}=3.8(2)$), where the chemical potential $\mu>U$, we observe a wedding-cake structure, where metallic, MI, and band-insulating (BI) phases coexist. A BI core of doubly-occupied sites forms in the center, visible as an extended region of empty sites with low occupation variance. This region is surrounded by a MI with a small number of defects. At the interface of these two phases and at the outer edge of the MI there is a metallic phase with large occupation variance.  For increasing interaction strength the MI ring broadens owing to the increased energy cost for double occupation.  At the same time the BI region decreases in size, until it disappears entirely for the largest interaction, where $\mu<U$. Fig. \;1C shows radial profiles of the average detected occupation, $\overline{n_{\mathrm{det}}}(r)$, and variance, $\sigma^2_{\mathrm{det}}(r)$, obtained from single experimental images by exploiting the symmetry of the trap and taking azimuthal averages over equipotential regions \cite{supplementary}.

An essential requirement for a MI state is a temperature well below the energy gap $k_{\mathrm{B}}T\ll U$, where $k_{\mathrm{B}}$ denotes the Boltzmann constant. We perform thermometry of the atomic gas in the lattice by comparing the detected radial occupation and variance profiles obtained from single images to theoretical calculations based on a second-order high-temperature series expansion of the Fermi-Hubbard model \cite{Oitmaa2006}. The effect of the harmonic trap is taken into account using a local density approximation, where the chemical potential varies locally \cite{Scarola2009}. The temperature and chemical potential are obtained from a fit to the detected density distribution $\overline{n_{\mathrm{det}}}(r)$, and all other parameters are calibrated independently. We find excellent agreement with theory for all interactions and measure temperatures as low as $k_{\mathrm{B}}T/U = 0.05$ for the largest interaction, corresponding to an average entropy per particle of $S/N = 1.15\,k_{\mathrm{B}}$. For weaker interactions we find values as low as $S/N = 0.99\,k_{\mathrm{B}}$. 
When repeating the experiment with the same parameters, we find a shot-to-shot variance in entropy consistent with fit errors \cite{supplementary}. The agreement with theory shows that the entire system is well described by a thermally equilibrated state and the underlying trapping potential is well described by a harmonic trap. 

\begin{figure*}[tb]
\centering
    \includegraphics{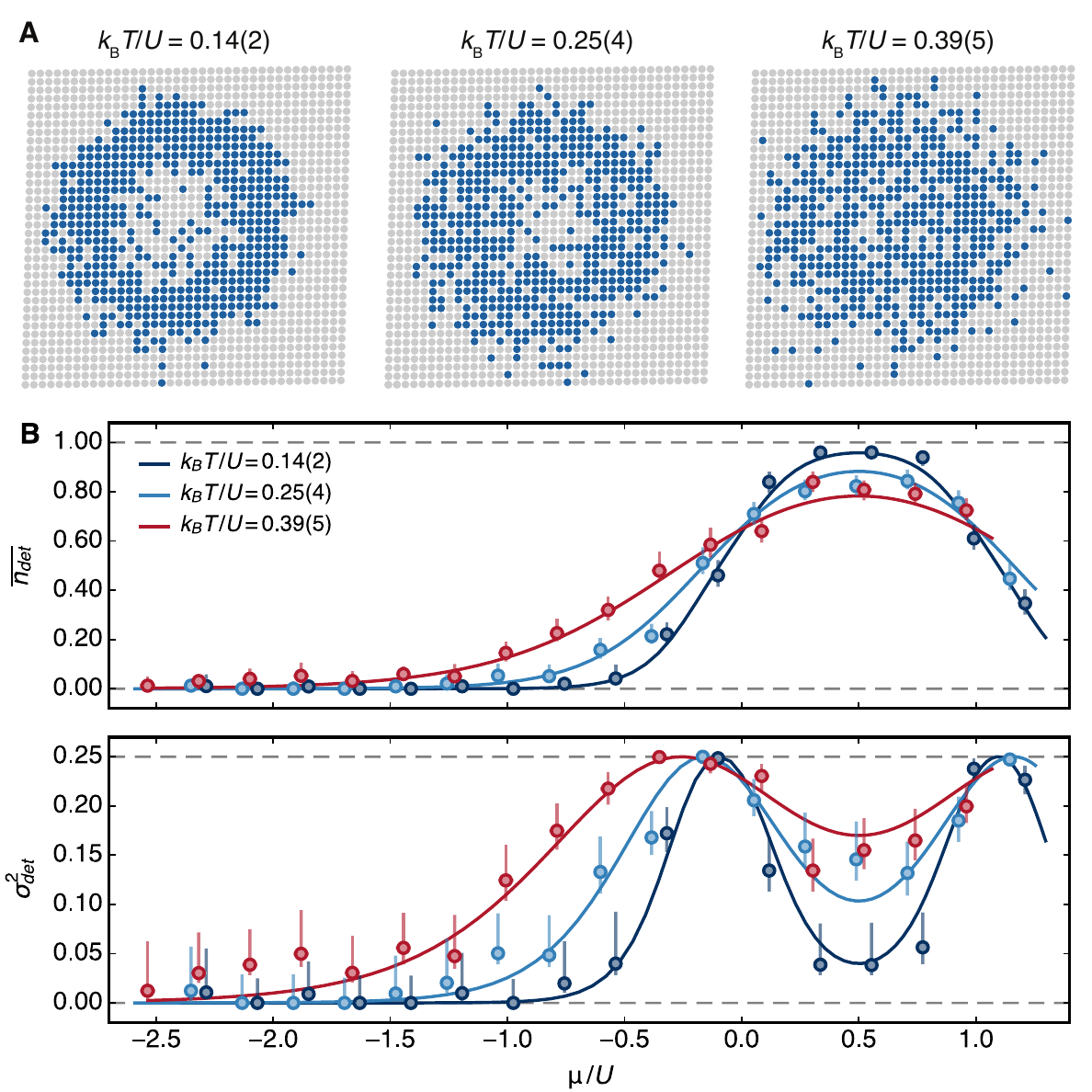}
    \caption{{\bf Melting a fermionic Mott insulator} (A) Detected site occupations from single images at different temperatures for $U/8\overline{t}=3.8(2)$. The clouds are heated by holding the atom cloud in a crossed dipole trap before lattice ramp-up. Hold times, from left to right, are $0\,\mathrm{s}$, $1\,\mathrm{s}$ and $3\,\mathrm{s}$. (B) Corresponding occupations and variance profiles versus chemical potential with error bars as in Fig. 1. The dark blue, light blue, and red curves are theory fits used to determine the temperature.}
	\label{fig4}
\end{figure*}

In Fig.\;2B we show $\overline{n_{\mathrm{det}}}$ and $\sigma^2_{\mathrm{det}}$ as a function of $\mu/U$, which corresponds to a scan along a single line in the $(\mu/U,k_{\mathrm{B}}T/U)$ phase diagram of the Hubbard model at $t\ll U$. These data are directly obtained from the respective radial profiles in Fig.\;1. The MI is identified by an extended region in $\mu/U$ with constant occupation $\overline{n_{\mathrm{det}}}=1$ and a strongly reduced variance. Because in this regime the total and detected fillings are approximately equal (see Fig.\,1B, right panel), the compressibility in the range just below half-filling ($\mu<U/2$) can be obtained from $\kappa = \partial n_{\mathrm{det}} /\partial \mu$, which is small in the MI region. The metallic regions are signaled by an enhanced compressibility with a peak in the variance distribution. 
The width of each peak is determined by the temperature and is smallest for the largest interactions, as $k_{\mathrm{B}}T/U$ decreases for increasing interactions in the experiment. 

Two contributions reduce the filling from one particle per site in the MI region: the finite temperature of the gas and the imaging fidelity. We determine a lower bound on the filling by averaging over $50$ images, resulting in $\overline{n_{\mathrm{det}}} = 96.5(2)\%$, limited by our finite imaging fidelity of $97.5(3)\%$. This filling gives an upper bound on the charge entropy (i.e. entropy involving density excitations in the atomic cloud) of $0.175,k_{\mathrm{B}}$ \cite{supplementary}. Because the fermionic particles are in two different spin states, the total entropy also contains a contribution from the spin entropy. From the fitted temperatures we calculate the total entropy per site as a function of $\mu/U$ for the different interactions (Fig.\;2B, inset). In the MI region we find $s_i = 0.70(1)\,k_{\mathrm{B}}$, consistent with $s_i = \,k_{\mathrm{B}}\,\mathrm{ln}\,2$, which corresponds to the entropy of a paramagnetic MI. In the BI region the entropy drops below $k_{\mathrm{B}}\,\mathrm{ln}\,2$, as in this case $78(1)\, \%$ of the lattice sites are occupied with doublons. In the BI the charge entropy can be estimated from the measured average density of excitations corresponding to sites with one particle. This gives $0.53(2)\,k_{\mathrm{B}}$, in good agreement with the calculated value of $0.5(1)\,k_{\mathrm{B}}$ from the theory fit. The variation in local entropy over the atomic cloud indicates that there is not only mass transport, but also efficient transport of entropy over many sites when loading the atoms into the lattice. This entropy transport could be the starting point for generating other low-temperature phases of matter using entropy redistribution or other cooling schemes \cite{Bernier2009, Ho2009, Lubasch2011, Bakr2011a}.

An accurate experimental study of the low-temperature Hubbard model generally requires large system sizes. By adjusting the evaporation we experimentally control the size of the MI with detected total atom numbers (without doublons) ranging from $N=154$ to $N=718$ (Fig.\;3). In all cases we find high values for the detected central occupation $\overline{n_{\mathrm{det}}}>0.92$ \cite{supplementary}. To investigate the reproducibility we repeat the same experiment and find that the standard deviation corresponds to $<5\%$ of the mean atom number. 

Whereas for temperatures $k_{\mathrm{B}}T\ll U$ a large MI with a sharp occupation distribution appears, the insulator is expected to gradually melt with increasing temperatures and eventually disappear. We experimentally control the temperature while keeping the atom number approximately constant by adjusting the evaporation and preparation of the atomic cloud \cite{supplementary}. We observe a clear change in the occupation and variance distribution from single images, which is also apparent in the respective distributions as a function of $\mu/U$ (Fig.\;4). For higher temperatures the occupation distribution broadens significantly, whereas the variance flattens and is only weakly suppressed at the highest detected occupations ($\mu=U/2$). The temperatures determined from theory comparison are in the range $k_{\mathrm{B}}T/U=0.14-0.55$, which corresponds to entropies of $S/N=0.97\,k_{\mathrm{B}}-1.58\,k_{\mathrm{B}}$. This shows that temperatures $k_{\mathrm{B}}T \ll U$ are required for MI states. 

For the lowest detected entropy value $S/N=0.97\,k_{\mathrm{B}}$ we expect strong antiferromagnetic correlations on nearest-neighbor sites \cite{Fuchs2011, Tang2012, LeBlanc2013, Imriska2015}. This should be detectable in single images by measuring the spin correlation function via spin-sensitive imaging. Additionally, our experiment is well suited for studying the competition of ferromagnetic and antiferromagnetic domains in the regime of weak lattices and large scattering lengths \cite{Ma2012}, investigating the phase diagram of polarized samples with repulsive or attractive interactions \cite{Snoek2008, Wunsch2010, Koetsier2010, Gukelberger2015}, and testing various entropy distribution schemes \cite{Bernier2009, Ho2009, Lubasch2011}.

\setlength{\parindent}{0pt}

\textbf{Acknowledgements} 
We acknowledge insightful discussions with Immanuel Bloch, Susannah Dickerson, Manuel Endres, Gregor Jotzu and Adam Kaufman. We acknowledge support from ARO DARPA OLE, AFOSR MURI, ONR DURIP, and NSF. D. G. acknowledges support from Harvard Quantum Optics Center and SNF. M.P, A.M, C.C acknowlege support from the NSF GRFP. 

\clearpage
\newgeometry{a4paper,
left=0.75in,right=0.483in,top=0.75in,bottom=1.55in
}

\onecolumngrid
~ ~
\begin{center} 
\begin{Large} \textbf{Supplementary Material} \end{Large}
\end{center}
\medskip
\smallskip
\twocolumngrid

\setcounter{section}{0}
\setcounter{subsection}{0}
\setcounter{figure}{0}
\setcounter{equation}{0}
\setcounter{NAT@ctr}{0}

\renewcommand{\figurename}[1]{FIG. }
\renewcommand{\theequation}{S\arabic{equation}}

\makeatletter
\renewcommand{\bibnumfmt}[1]{[S#1]}
\renewcommand{\citenumfont}[1]{S#1}
\renewcommand{\thefigure}{S\@arabic\c@figure} 
\makeatother
\renewcommand{\theequation}{S\arabic{equation}} 
\renewcommand\thetable{S\arabic{table}}
\renewcommand{\vec}[1]{{\boldsymbol{#1}}}
\renewcommand\Re{\operatorname{Re}}
\renewcommand\Im{\operatorname{Im}}

\section{Experimental sequence}

\subsection{Cloud preparation and evaporative cooling}

\begin{figure}[tb]
\centering
    \includegraphics[width=\columnwidth]{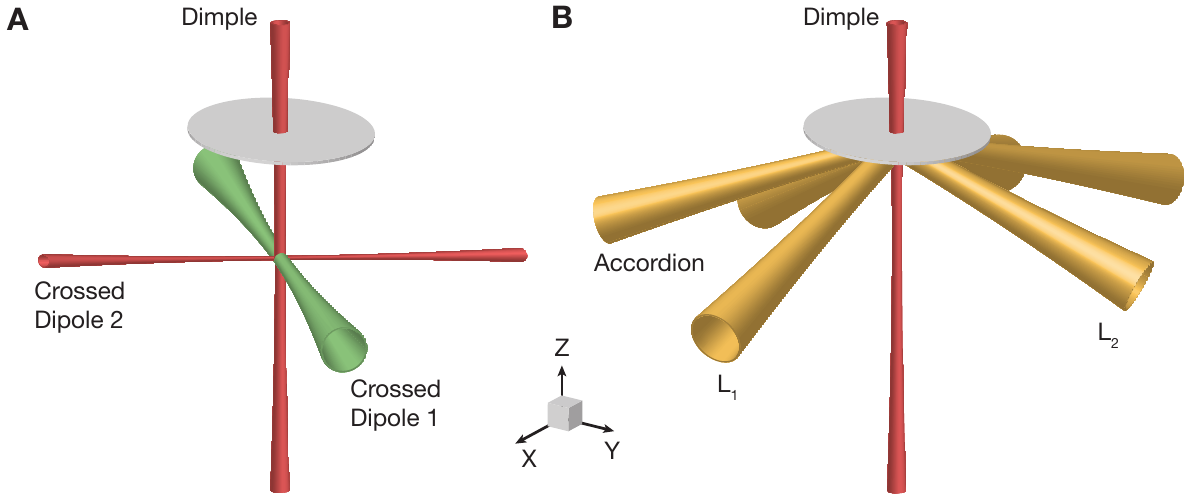}
    \caption{Geometry of laser beams used in the experiment. The dimple trap enters the glass cell through the objective and substrate (grey disk), and is shown with the substrate in both subfigures as a reference. (A) The ``crossed dipole 1" beam is used for transport and propagates along the long axis of the glass cell, below the substrate. The ``crossed dipole 1" beam intersects ``crossed dipole 2" at approximately $65^{\circ}$. (B) The accordion and square optical lattice beams reflect off of the substrate, and the lattice beams are additionally retroreflected using curved mirrors (not shown). Laser beam wavelengths are indicated through the colors shown (red: $780$\,nm, yellow: $1064$\,nm, green: $1070$\,nm). Beam waists and distances are not to scale.}
	\label{sifig1}
\end{figure}

In the following we describe the experimental protocol for creating a two-dimensional Fermi gas at a distance of $10\,\mu\mathrm{m}$ from a superpolished substrate. After initial laser cooling, optical pumping, and evaporative cooling steps, a balanced spin mixture of about $10^7$ atoms in the two lowest hyperfine states $|1 \rangle$ and $|2 \rangle$ of the $2^2\mathrm{S}_{1/2}$ electronic ground state is transported from the MOT chamber to the glass cell, $100\,\mu\mathrm{m}$ below the superpolished substrate (see Fig. \ref{sifig1}A). To transport the atoms we move the focus of a red-detuned laser beam (crossed dipole 1) at $\lambda=1070$\,nm and a depth of $100\,\mu \mathrm{K}$ by translating a lens on an air-bearing stage. After transport we set a magnetic bias field of $300$\,G along the $z$-direction ($a=-279\,a_0$). We then load the atoms into a crossed dipole trap by turning on an additional red-detuned laser beam (crossed dipole 2) at wavelength $\lambda = 785$\,nm, which is derived from a superluminescent LED (sLED, Exalos EXS0800-025-10-0204131) and amplified by a tapered amplifier. In this configuration we perform a second evaporative cooling step by first lowering the powers of both beams and then turning on a magnetic field gradient of $60$\,G/cm along the vertical direction. This reduces the bias field at the position of the atoms to about $176$\,G ($a=-197\,a_0$) and tilts the trapping potential, allowing hot atoms to escape. 

After removing the magnetic field gradient we load the atoms into a single layer of a vertical lattice with $40\,\mu \mathrm{K}$ depth and $15\,\mu\mathrm{m}$ spacing along the $z$-direction, which is formed by reflecting a $1064$\,nm beam under an angle of $2^{\circ}$ from the superpolished substrate. We control the spacing of this ``accordion" lattice by varying the reflection angle from the substrate via a galvanometer controlled mirror. To provide an additional controllable harmonic confinement in the $x-y$-plane, we add another red-detuned sLED beam at $785$\,nm, which propagates through the microscope objective along the $z$-direction (dimple trap). After loading the atoms into the ``accordion" lattice and dimple trap, we transport the atoms to a distance of $10\,\mu\mathrm{m}$ below the superpolished substrate by increasing the reflection angle to $17^{\circ}$. In this configuration the atomic cloud has an aspect ratio of 1:440:130. Final evaporation is performed by lowering the depth of the dimple trap in the presence of a strong magnetic field gradient of $64$\,G/cm along the horizontal direction, maintaining a bias field of $243$\,G ($a=-262\,a_0$). By varying the endpoint of this evaporation, we prepare samples with a controllable atom number. Finally, the magnetic gradient is removed, the harmonic confinement from the dimple trap is reduced, and the magnetic bias field is increased to the values used in the experiment $B=\{538, 550, 560, 620\}$\,G, corresponding to repulsive scattering lengths $a=\{37, 85, 129, 515\}\,a_0$ \cite{Zurn2013-SI}. The highest value of $620$\,G is chosen such that little additional heating due to the Feshbach resonance is observed for our densities. For all magnetic bias fields except the lowest value we quickly ramp the field across the narrow s-wave resonance located at $543$\,G to avoid heating and losses \cite{Strecker2003-SI}. This is done by rapidly switching on an additional magnetic field of about $5$\,G within $3\,\mu\mathrm{s}$, by discharging a capacitor charged to $100$\,V through an additional coil.

To prepare samples with different atom numbers (data in Fig. 3 of the main manuscript) we adjust the final evaporation power of the dimple beam. For the data in Fig. 4 of the main manuscript, where we change the temperature of the gas before loading it into the lattice, we add a hold time ranging between $0-3$\,s after the evaporation but before loading the atoms into the lattice. Owing to underlying heating mechanisms (e.g. proximity to a Feshbach resonance or scattering of dipole trap photons), the gas heats up while the atom number is reduced by about $15\%$ for $3$\,s as compared to $0$\,s.

\subsection{Optical lattice}

After the preparation we load the Fermi gas into a square optical lattice, formed by two red-detuned and retro-reflected laser beams along the $x$ and $y$ directions at a wavelength of $\lambda=1064$\,nm, denoted in Fig. \ref{sifig1}B as L$_1$ and L$_2$. The lattice are additionally reflected under an angle of $21^{\circ}$ from the surface of the substrate. We use a $30$\,ms long linear ramp of the two laser beam powers while at the same time turning off the ``accordion" lattice and the dimple trap. In the final configuration the expression for the two-dimensional square lattice potential of the atoms in the $x-y$ plane is given by
\begin{equation}
V(x,y) = V_x \cos^2(\pi x/l) + V_y \cos^2(\pi y/l).
\end{equation}
We set the lattice depths to $V_x=12.5(2)\,E_{\mathrm{R}}$ and $V_y=15.9(2)\,E_{\mathrm{R}}$, where $E_{\mathrm{R}}=h^2/(8ml^2)$ denotes the recoil energy, $m$ is the atomic mass of $^6\mathrm{Li}$, and $l=0.569\,\mu\mathrm{m}$ is the lattice spacing. The atoms are trapped in a single layer of the $1.5\,\mu\mathrm{m}$-spacing lattice along the $z$-direction, which is created by the additional reflection of the lattice beams from the superpolished substrate. In all pictures in the experiment we observe no evidence of atoms in different planes. The harmonic trapping frequency along the $z$ direction is $\omega_z/2\pi=95$\,kHz, which is sufficiently large to ensure that the Fermi gas is well within the two-dimensional regime with $\mu/\hbar\omega_z < 0.11$ and $k_{\mathrm{B}}T/\hbar\omega_z < 0.02$. The presence of atoms in a single layer of the lattice is ensured by loading from a two-dimensional dipole trap, where $\mu/\hbar\omega_z < 1.0$ and $k_{\mathrm{B}}T/\hbar\omega_z < 0.22$.

At the same time the largest scattering length is below $4\%$ of the harmonic oscillator length along the $z$-direction. The underlying harmonic trap frequency in the $x-y$ plane due to the finite waists of the lattice beams are $\omega_x/2\pi=0.690(9)$\,kHz and $\omega_y/2\pi=0.604(1)$\,kHz.

\begin{figure*}[tb]
\centering
    \includegraphics[width=0.7\linewidth]{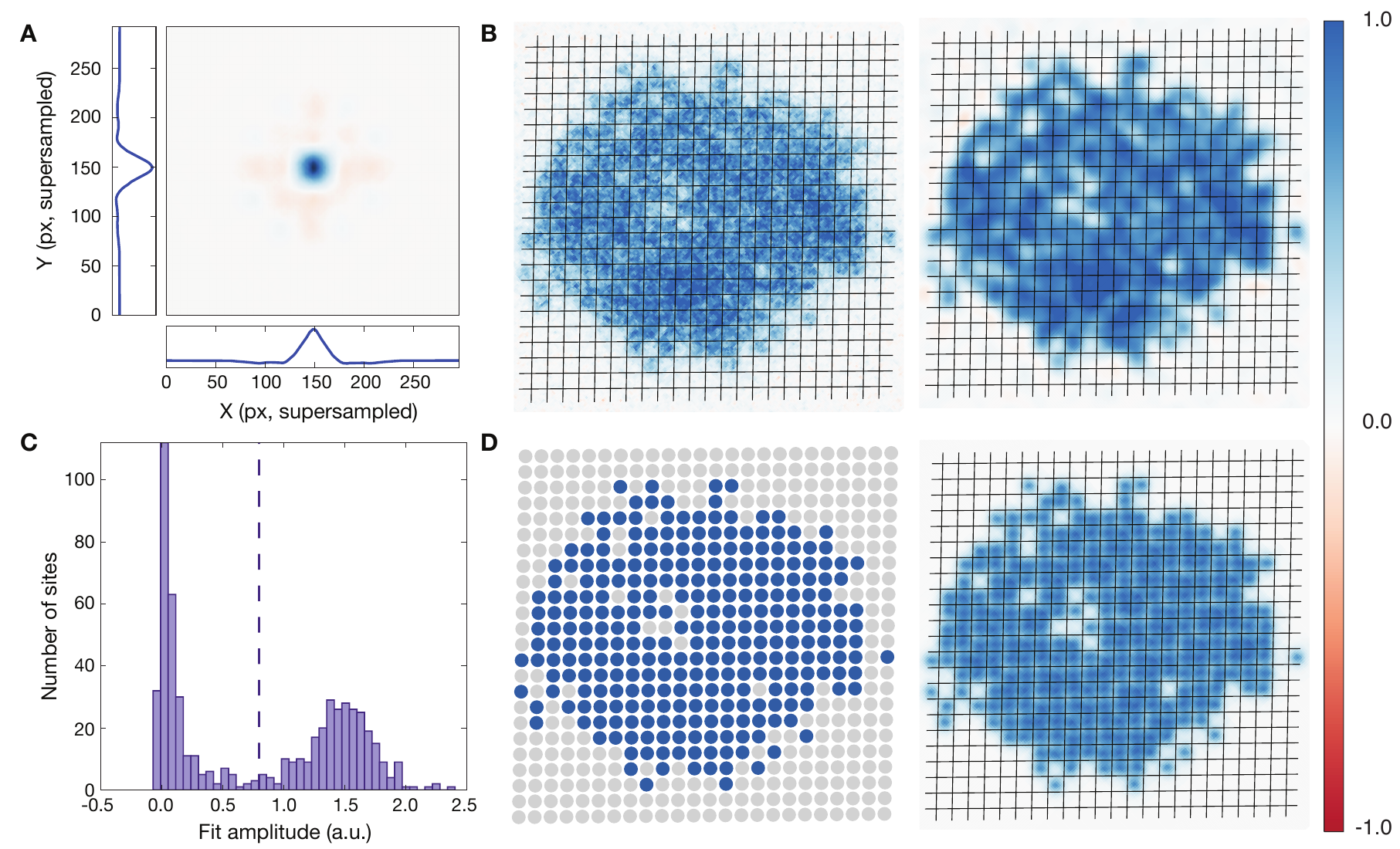}
    \caption{(A) PSF-derived kernel used in image deconvolution. The plot margins show cuts through the kernel. (B) Image before and after deconvolution with the kernel, with lattice geometry overlaid. (C) Histogram of overlap amplitudes on each site in arbitrary units, for the convolved image. The chosen threshold value is indicated by the vertical line. (D) Resulting binarization of the image based on the threshold value, indicated by filled circles on each occupied lattice site (left), as well as point spread functions on each occupied lattice site (right). The colorbar (arbitrary units) for (A), (B), and (D) is shown on the far right, symmetric about zero.}
	\label{sifig2}
\end{figure*}

\subsection{Imaging}

Our imaging sequence begins by freezing the atom distribution, ramping up the square optical lattice to depths of  $V_x=50.2(7)\,E_{\mathrm{R}}$ and $V_y=53.8(5)\,E_{\mathrm{R}}$ in $1$\,ms, then up to the depths and trap frequencies required for imaging in $100$\,ms. An additional lattice along the $z$ direction, also at a wavelength of $\lambda=1064$\,nm, is ramped up concurrently and in the same manner, to provide added confinement in the $z$ direction. This allows us to achieve nearly-degenerate on-site trap frequencies of $(\omega^{\mathrm{site}}_x, \omega^{\mathrm{site}}_y, \omega^{\mathrm{site}}_z) = 2 \pi \times (1.46, 1.45, 1.47)$\,MHz. At this point, we apply the Raman sideband imaging scheme described in \cite{Parsons2015-SI}. Our imaging technique is not sensitive on the spin and measures the site-resolved occupation, which is either one for a single particle and zero for an empty or doubly-occupied site. 

\section{Site-resolved imaging and reconstruction algorithm}

We implement a deconvolution-based image analysis technique to reconstruct the atom distribution. In this technique, a kernel associated with a given site is generated to have minimum overlap with the signal from atoms on neighboring sites, and unity overlap with the signal from an atom on the same site. The convolution of the kernel with an image then allows extraction of signal amplitudes on each site. From a histogram of extracted signal amplitudes and a threshold we then determine which sites were occupied. This method gives the same results as directly fitting the amplitudes of point spread functions (PSFs) on every site, but is significantly faster in computation. 

The technique relies on the existence of a kernel such that the overlap of the kernel $k_{ij}(\vec{r})$ on site $(i, j)$ with a PSF $w_{l,m}$ on site $(l,m)$ is given by
\begin{equation}
\int k_{i,j}(\vec{r}) w_{l,m}(\vec{r}) d\vec{r} = \delta_{i,l} \delta_{j,m}. 
\end{equation}
Across our imaging field of view, there is negligible distortion. As such, the PSFs from each site are identical; $w_{ij}(\vec{r}) = w(\vec{r}-\vec{r_{ij}})$ for the PSF on site $(i, j)$, where $w(\vec{r})$ is the measured average PSF. Because of this, it is natural to express the kernel $k(\vec{r})$ as a linear combination of the point spread functions $w_{ij}(\vec{r})$. It also then follows that $k_{ij} = k(\vec{r}-\vec{r_{ij}})$.

As the width of the measured PSF is small relative to the lattice spacing, we find that we can approximate our kernel by one generated using only PSFs over a five by five site region:
\begin{equation}
k(\vec{r}) = \sum_{i, j \in [-2, 2]} b_{ij} w_{ij}(\vec{r})
\end{equation}
We then find the $b_{ij}$ by minimizing the overlap of $k(\vec{r})$ with every PSF $w_{ij}(\vec{r})$ of the five-by-five grid, except for the central one at $(i,j)=(0,0)$. The overall normalization is decided such that the overlap of $k(\vec{r})$ with $w_{00}(\vec{r})$ is unity.

The kernel is generated once from a measured point spread function and lattice geometry. The PSF is determined by averaging several isolated peaks, taken from sparsely-populated images with centers aligned with sub-pixel resolution \cite{Parsons2015-SI}. The lattice geometry is determined from a Fourier transform of an averaged image, where the peaks in Fourier space correspond to the lattice periodicity. The resulting kernel is shown in Fig. \ref{sifig2}A.

By placing the kernel on any given lattice site and computing its overlap with the image, we find the amplitude of that site. If this is done for every lattice site, it is equivalent to taking a discrete deconvolution of step size equal to the lattice spacing. However, because we have multiple pixels per lattice site and our lattices are not well-aligned to our pixel array, we instead take a discrete deconvolution of step size equal to one pixel. This yields an image (Fig. \ref{sifig2}B) still containing multiple pixels per lattice site, but with the amplitudes given by the pixel values at the center of each lattice site.

Between individual images, we see lattice translations on the order of a few percent of a lattice site. To account for this, we take the Fourier transform of each individual image and use the phase information at the lattice peaks to determine the offset for that image. The final step is to create a histogram of the amplitudes on each site (Fig. \ref{sifig2}C) and apply a threshold to assign whether each site is occupied (Fig. \ref{sifig2}D).

\begin{figure}[tb]
\centering
    \includegraphics{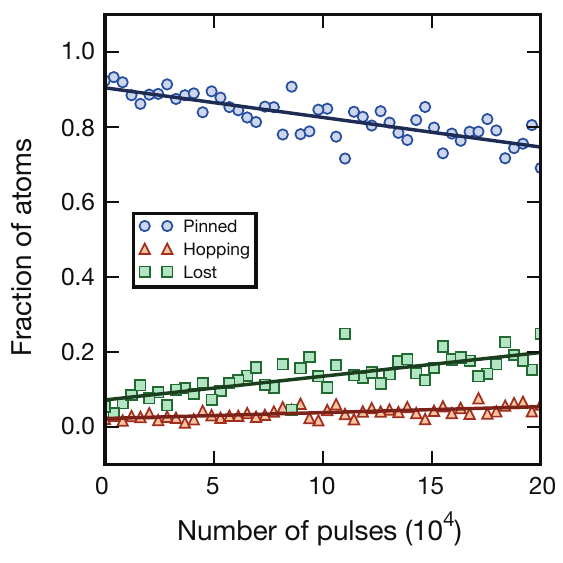}
    \caption{Fraction of atoms that are pinned, lost, or hop in a second imaging frame relative to the first, for a varying number of additional pulses in between the two frames (x-axis). The region of interest is a square of $50 \times 50$ sites around a cloud with a radius of about $12$ sites. The average density within the cloud is $85\%$. Lines are fits to the data.}
	\label{sifig3}
\end{figure}

By taking two images of a single densely-populated cloud with a varied number of additional pulses in between, we can determine the fraction of atoms that are pinned, hopping, or lost ($f_p$, $f_h$, or $f_l$) after a single image. The images themselves are taken with $6.4 \times 10^4$ imaging pulses, consistent with the number of pulses used for the data presented in the main text. The fits shown in Fig. \ref{sifig3} have the following functional forms, for number of pulses $n$:

\begin{align}
f_p(n) &= 0.905(10) - [7.9(8) \times 10^{-7}]n\\
f_h(n) &= 0.023(3) + [(1.6(3) \times 10^{-7}]n\\
f_l(n) &= 0.072(9) + [(6.3(8) \times 10^{-7}]n
\end{align}

This yields $f_p = 94.9(5)\%$, $f_h = 1.0(2)\%$, and $f_l = 4.1(5)\%$ between the two images. The histogram and threshold indicate that atoms which hop in the second half of an imaging sequence will still be counted by our reconstruction algorithm. Then, the probability of correctly determining the occupation of a site $F$ can be estimated to be $F = 1 - (m_l - m_h)n/2$, where $m_l$ ($m_h$) are the fitted rates of fractional atom loss (hopping) per pulse \cite{Parsons2015-SI}. From this we obtain $F = 97.5(3)\%$.

\section{Theoretical model}

\subsection{High-temperature series}

The experiment is well described by the single-band Fermi-Hubbard model in the grand-canonical ensemble in the presence of an additional harmonic trap 
\begin{equation}
    \hat H   =  -\sum_{\langle ij\rangle,\sigma}t_{ij}(\hat c^\dagger_{i\sigma}\hat c_{j\sigma}+\mathrm{h.c.}) 
    + U\sum_i \hat n_{i\uparrow}\hat n_{i\downarrow} + \sum_{i,\sigma} (V_i-\mu) \hat n_{i\sigma}\,.
\end{equation}
Here $\hat c^\dagger_{i\sigma}$ and $\hat c_{i\sigma}$ denote the fermionic creation and annihilation operators for the two spin states $\sigma\in\{\uparrow,\downarrow\}$, the density operator is denoted by $\hat n_{i\sigma}=\hat c^\dagger_{i\sigma} \hat c_{i\sigma}$ and $V_i=0.5ml^2(\omega_x^2i_x^2+\omega_y^2i_y^2)$ is the trap energy with lattice site indices $i_x$ and $i_y$ along the lattice directions. 
The Hubbard on-site interaction is denoted by $U$, the chemical potential is $\mu$ and the tunneling is $t_{i,j}=t_{x}$ ($t_{i,j}=t_{y}$) along the $x$ ($y$) lattice directions. 
We include the effect of the harmonic trap in a local density approximation (LDA) by considering a local chemical potential varying quadratically with distance to the trap center $\mu_i = \mu-V_i$.
It is hence sufficient to calculate all relevant quantities for the homogeneous Fermi Hubbard model in the grand-canonical potential. 
We use a high-temperature series expansion up to second order of the partition function, where the expansion parameters is given by $k_{\mathrm{B}}/t$ \cite{Oitmaa2006-SI}. This approach is expected to be accurate if $t_{x,y} < k_{\mathrm{B}}T < U$. From the partition function we obtain a second-order expression of the grand-canonical potential per site
\begin{equation}
-\beta \Omega = \mbox{ln}(z_0) + \beta^2 (t_x^2+t_y^2) \frac{1}{z_0^2} \left(2\zeta(1+\zeta^2 w)+\frac{4\zeta^2}{\beta U}(1-w)\right), 
\end{equation}
where $z_0 = 1+2\zeta +\zeta ^2w$ is the single-site partition function. Here we use the abbreviations $\beta=1/k_{\mathrm{B}}T$, $\zeta=\mathrm{exp(\beta \mu)}$ and $w=\mathrm{exp(-\beta U)}$. 
Thermodynamic quantities such as the single-site density and entropy can be obtained from derivatives with $n=-\partial \Omega / \partial \mu$ and $s=-\partial \Omega / \partial T$, while the probability for a doubly occupied site is $p_d = \partial \Omega / \partial U$.
The probability for a single particle of any spin per lattice site is then
\begin{equation}
p_s = n - 2p_d\,.
\end{equation}
This quantity corresponds to the experimentally detected single-site occupation $\overline{n_{\mathrm{det}}}$. 

We fit the experimentally measured radial site occupation distributions to theory with $T$ and $\mu$ as the two free fit parameters, while all other parameters are fixed and calibrated independently. From this we calculate the total atom number and entropy by integrating over the entire system.

\subsection{Entropy estimates}

The charge entropy per site (i.e. entropy involving density excitations) in the MI regime can be obtained from the expression 
\begin{equation}
s^{\mathrm{charge}}_i=-k_{\mathrm{B}}\sum_j p_j \ln (p_j). 
\label{eq-chargeentropy}
\end{equation}
Here $p_j$ denotes the occupation probability in the grand-canonical ensemble of each possible quantum state $j$ on lattice site $i$. Neglecting the spin degree of freedom, there are three possibilities: an empty site ($p_h$), a doubly occupied site ($p_d$), or a single atom ($p_s$). By averaging over several experiments we determine the average fraction of excitations in the Mott-insulator corresponding to holes or doublons $p_h+p_d = 1 - \overline{n_{\mathrm{det}}}$. As the imaging does not distinguish between an empty and doubly-occupied site, we provide an upper and lower bound for the charge entropy by either setting the hole and doublon probabilities equal or by setting one of them to zero. In both cases the general dependence of the charge entropy on $p_s$ (divided by ${\mathrm{B}}$ is then similar to that of the single site variance, which is given by $p_s(1-p_s)$.

For the band-insulating region we perform a similar calculation. In this case most of the sites are occupied with a doublon and a few sites show excitations in form of single particles. In this case we neglect contributions from empty sites, as these require significantly higher excitation energies. 

\subsection{Mass and entropy redistribution during lattice loading}

To achieve the entropy and density profiles shown in figures 1 and 2 of the main text when loading the lattice, the entropy and mass of the atomic cloud must redistribute substantially as the lattice is turned on.  We can see this by comparing the local density, $n(r)$, and the local entropy density, $s(r)$, obtained from the fitted occupation profiles in the lattice with calculated profiles for an ideal two-dimensional Fermi gas in the harmonic trap just before lattice loading.  To compute the thermodynamic quantities for the harmonically trapped two-dimensional gas we apply a local density approximation, letting the chemical potential vary according to 

\begin{equation}
\mu(r) = \mu_{0} - \frac{1}{2} m \omega^{2} r^{2},
\label{eq-lda}
\end{equation}

where $\mu_{0}$ is the chemical potential in the center of the cloud, $m$ the atomic mass, and $\omega$ the trap frequency. Following the procedure in \cite{Landau5-SI} for obtaining the thermodynamic potential of a homogeneous three-dimensional fermi gas, we compute the local thermodynamic potential, $\Omega(r)$, of a two-dimensional Fermi gas under the local density approximation:

\begin{equation}
\Omega(r) = \frac{g V}{\beta^{2}} \frac{m}{2 \pi \hbar^{2}} \mathrm{Li}_{2}(-e^{\beta \mu(r)})
\label{eq-thermpot}
\end{equation}

Here, $g$ is the number of spin states, $V$ is the volume of the system, $\beta = 1/k_{B} T$, $T$ is the temperature, $\hbar$ is the reduced Planck constant, and $\mathrm{Li}_{2}()$ is the second-order polylogarithm function.  From the thermodynamic potential we can compute the density and entropy density:

\begin{equation}
n(r) = - \frac{1}{V} \left( \frac{\partial \Omega(r)}{\partial \mu}\right)_{T, V} = -\frac{g m}{2 \pi \beta \hbar^{2}} \ln(1 + e^{\beta \mu(r)})
\label{eq-density}
\end{equation}

\begin{eqnarray}
s(r) &=&  - \frac{1}{V} \left( \frac{\partial \Omega(r)}{\partial T}\right)_{\mu, V} \\ &=& -\frac{g m}{2 \pi \beta \hbar^{2}} \left[ \mu(r) \ln(1 + e^{\beta \mu(r)}) + \frac{2}{\beta} \mathrm{Li}_{2}(-e^{\beta \mu(r)}) \right] \nonumber
\label{eq-entropy}
\end{eqnarray}

\begin{figure}[h]
\centering
    \includegraphics[width=\columnwidth]{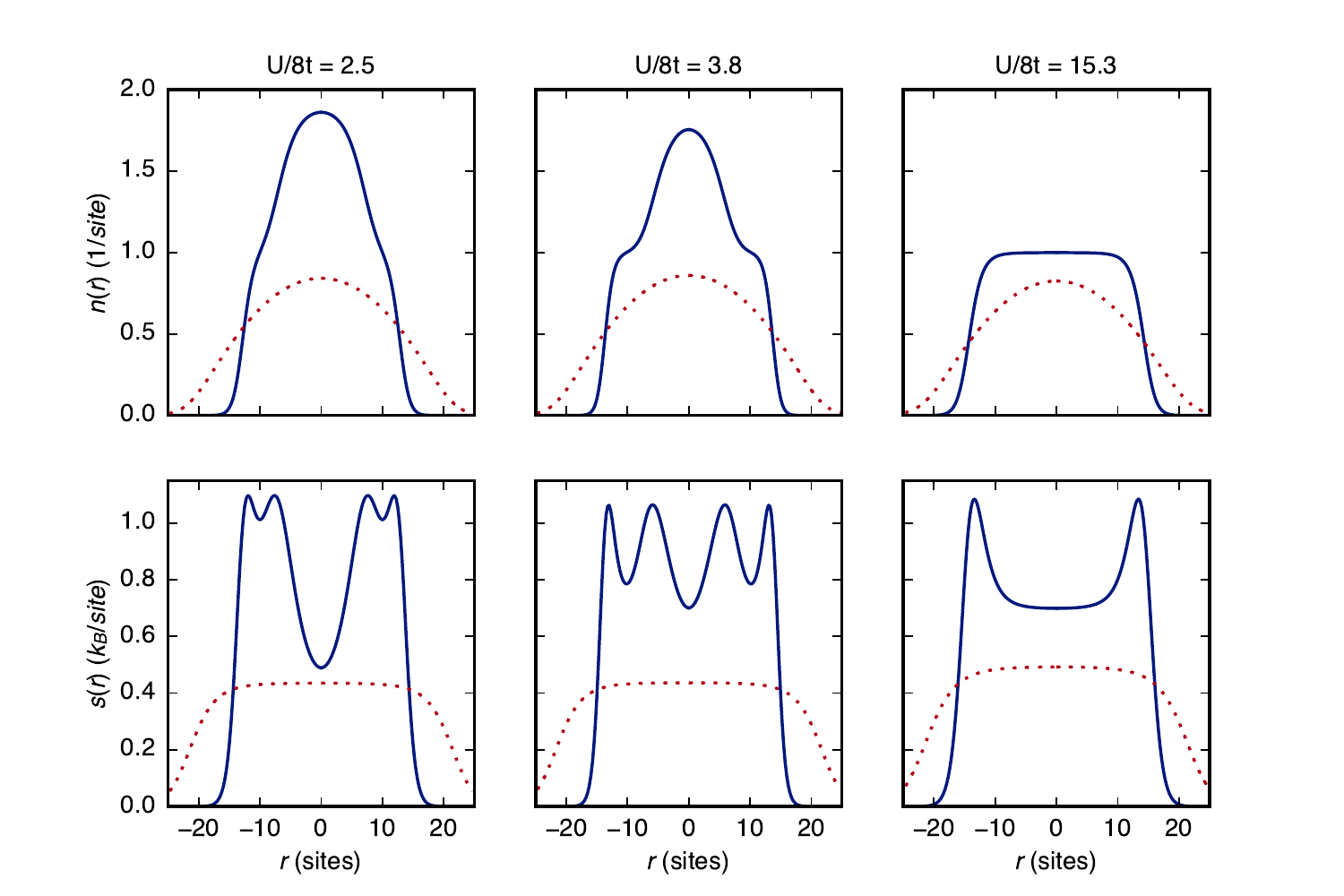}
    \caption{Comparison of mass and entropy distributions before and after the optical lattice is loaded.}
	\label{sifig5}
\end{figure}

The parameters $\mu_{0}$ and $T$ of the harmonically trapped gas are determined by matching the total entropy and atom number to the measured values in the lattice.  The harmonic trap frequency in the optical dipole trap before the lattice loading is $\omega = 2 \pi \cdot 562 \, \mathrm{Hz}$.  Using these equations and parameters we estimate the local density and entropy profiles for the atomic cloud just before that lattice is loaded (Fig. \ref{sifig5}, red dotted lines).  By comparing these profiles to the in-lattice profiles (blue solid lines), obtained from fits to a high-temperature series expansion as previously described, we can see that there is significant mass and entropy redistribution as the lattice is loaded.

\section{Radial Fits}

We obtain radial occupation and variance profiles $\overline{n_{\mathrm{det}}}(r)$ and $\sigma^2_{\mathrm{det}}(r)$, from individual binarized images by binning the lattice sites in elliptical annuli containing the same number of sites and calculating the mean and variance of the site occupations in each bin. We take the center of mass of each image as the center for azimuthal averaging. The principal axes of the ellipse are along the lattice directions. The aspect ratio is determined from the ratio of standard deviations of the spatial distribution along each principal axis, averaged over many images. The error bars on the mean and variance are 68\% confidence intervals determined from numerically computed sampling distributions of the mean and variance of a Bernoulli distribution whose mean is given by the measured mean occupation of each bin. We then fit $\overline{n_{\mathrm{det}}}(r)$ to the model described in the previous section. The error bars on reported temperatures and entropies are derived from the standard errors of the fit. We find that the standard deviation of the fitted entropy over many experimental cycles is consistent with standard errors from the fit.
To determine an estimate for the number of atoms in the MI region, we count all atoms within the central elliptical region where the fit gives a filling larger than $0.9$. For the data in Fig. 3 we determine the filling in the MI by averaging the measured site occupation in the central region of the atomic cloud over many experimental realizations. Tab. \ref{tab:fillings} summarizes the results. 

\begin{table}[ht]
\centering
\begin{tabular}{c|c|c|c|c|c|c}
$\overline{N}$ & 136 & 198 & 244 & 389 & 541 & 685 \\ [0.5ex]
\hline                  
MI filling & 0.918 & 0.965 & 0.970 & 0.960 & 0.943 & 0.919\\
[1ex]
\end{tabular}
\caption{{\bf Average MI fillings} The detected filling in the MI is shown for various samples with total average atom number $\overline{N}$, which is obtained by averaging the occupation in the central region of the atomic cloud. The experimental data of Fig. 3 is used for this table.}
\label{tab:fillings}
\end{table}

\section{System Calibrations}

\subsection{$U$ Calibration}

\begin{figure}[t]
\centering
    \includegraphics[width=\columnwidth]{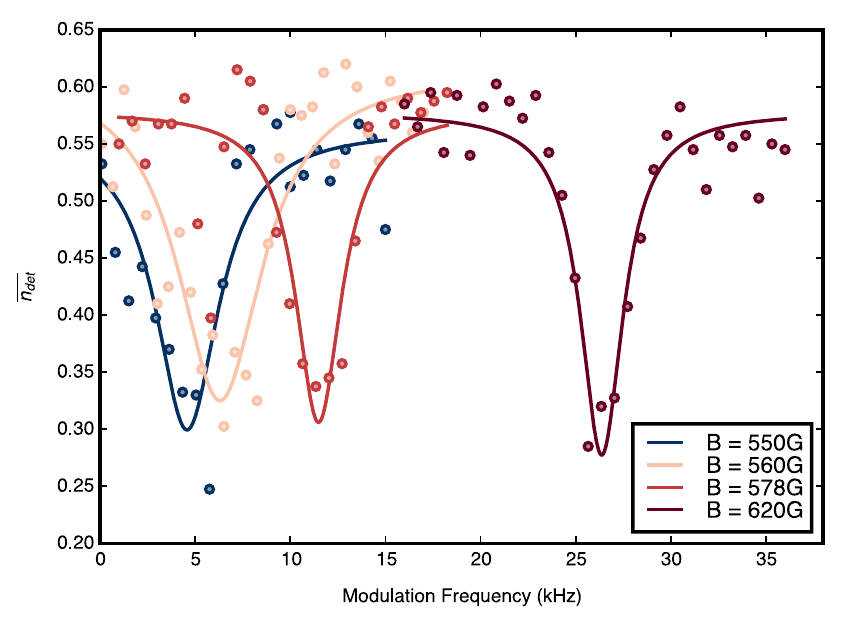}
    \caption{Amplitude modulation of one of the lattice beams at a frequency corresponding to the interaction energy, $\nu=U/h$, results in a decrease in the observed filling as doublons are created. The magnetic field is varied, which changes the scattering length and consequently the interaction energy.}
	\label{sifig4_u}
\end{figure}

The on-site interaction $U$ is calibrated using lattice modulation spectroscopy. A MI sample in the optical lattice is prepared in a $V_{x} = 12.5(2)\,E_{\mathrm{R}}$, $V_{y} = 15.9(2)\,E_{\mathrm{R}}$ deep lattice at $B= \{550, 560, 578, 620\}$\,G, corresponding to scattering lengths of $a=\{85, 129, 221, 515\}\,a_0$. The amplitude of the lattice along the $x$ direction is modulated sinusoidally at variable frequency $\nu$ with about $10\%$ modulation depth for $20$\,ms. If the modulation frequency matches the interaction energy, $\nu = U/h$, excitations in the form of doubly-occupied sites are produced, which reduces the observed density $\overline{n_{det}}$. For the measurement we average over a spatial region that includes the full cloud and some of the surrounding empty sites. Fig. \ref{sifig4_u} shows the averaged density as a function of modulation frequency for several interaction strengths. The positions of the resonances are identified from a Lorentzian fit. 

\begin{figure}[b]
\centering
	\includegraphics[width=\columnwidth]{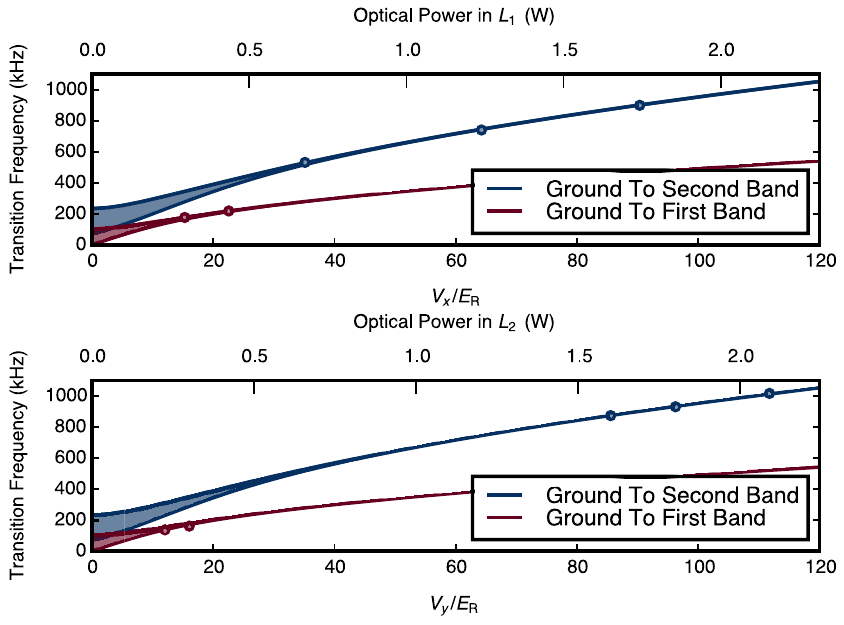}
	\caption{Amplitude modulation of lattice depth along the $x$ and $y$ directions at the transition frequency from the ground to second excited band results in a clear heating signal. Fitting the measured resonant frequencies to transition frequencies calculated from the band structure allows calibration of the lattice depth. Error bars are smaller than the points shown.}
	\label{sifig4_txty}
\end{figure}

\subsection{Lattice Depth Calibration}

To calibrate the lattice depth we apply lattice modulation spectroscopy and measure the transition frequencies between different energy bands. The atoms are prepared in a non-interacting state and loaded into either the lattice along the $x$ or $y$ direction. We increase the lattice to a variable power in $50$\,ms and then modulate at variable frequency $\nu$ with approximately $5\%$ modulation depth for $2000$ cycles. After the modulation, the lattice is ramped down to a constant depth and held for $10$\,ms before imaging. In the measured width of the real-space distribution we find resonances corresponding to the transition from the ground to the second excited band. We use a Lorentzian fit to determine the resonant frequency. From a comparison of the resonances to a band structure calculation we then determine the lattice depth calibration. The results are shown in Fig. \ref{sifig4_txty}, where the measured resonance frequencies are seen to be in good agreement with the fitted model.

While in a deep lattice the transition from the ground to the first excited band due to lattice amplitude modulation is forbidden by symmetry, this is not the case for shallower lattices. To verify our calibration, we reduce the lattice depth to $V_{i}\approx 20 E_{\mathrm{R}}$ and drive transition to the first excited band, finding good agreement of the transition frequency with the calculated band structure. 

\begin{figure}[tb]
\centering
    \includegraphics[width=\columnwidth]{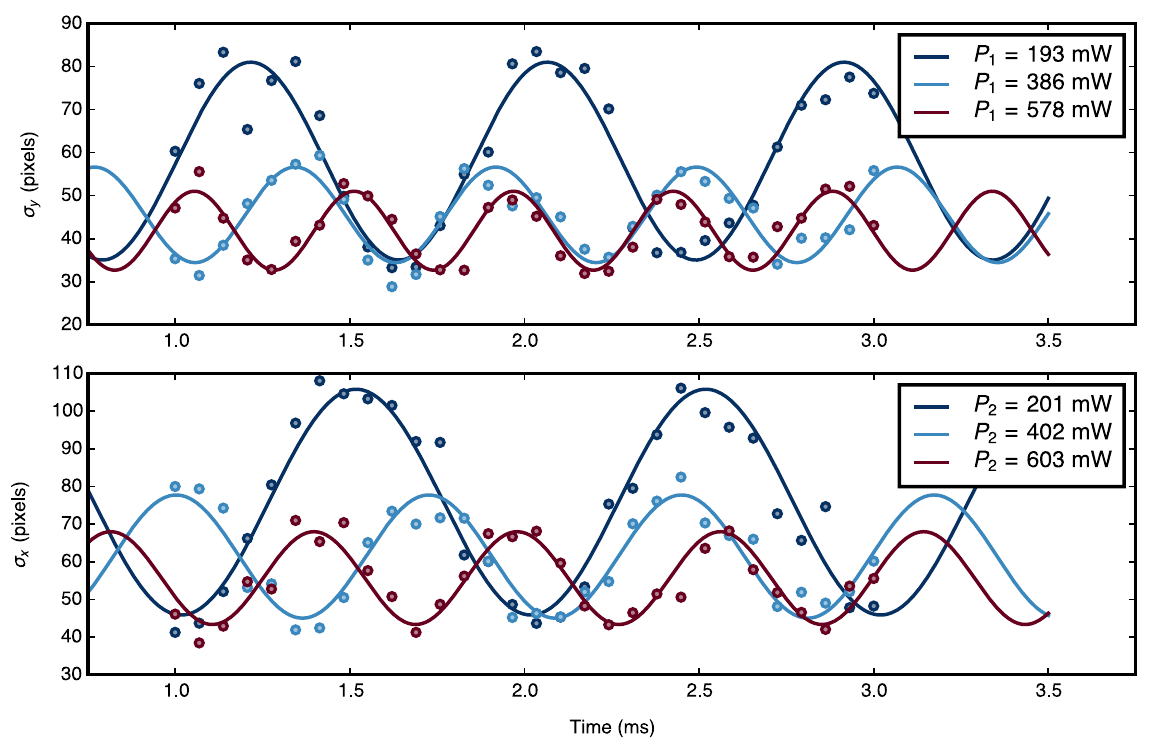}
    \caption{Breathing mode oscillations of atoms held by either L$_1$ or L$_2$. The standard deviation of the atomic fluorescence signal along the direction orthogonal to the lattice is shown as a function of time after a sudden relaxation of the harmonic confinement.}
	\label{sifig4_breathing}
\end{figure}

\subsection{Harmonic Trap Frequency Calibration}

The harmonic confinement due to the Gaussian profile of the lattice laser beams is calibrated by observing breathing mode oscillations. We load the atoms into a trap formed by the lattice of interest and the dimple beam. The dimple beam is suddenly switched off to excite the breathing mode along the direction normal to the lattice. After a variable oscillation time we pin the atomic distribution by ramping the lattice depths to $V_{x} = 50.2(7)\,E_{\mathrm{R}}$ and $V_{y} = 53.8(5)\,E_{\mathrm{R}}$ in $0.1$ ms and subsequently image the atomic distribution. A sinusoidal fit to the spatial width of the atomic distribution then yields twice the harmonic trap frequency, as shown in Fig. \ref{sifig4_breathing}.

\end{document}